\documentstyle[epsfig,prl,aps,12pt,tighten]{revtex}
\newcommand{\be}{\begin{equation}}
\newcommand{\ee}{\end{equation}}
\newcommand{\ba}{\begin{array}}
\newcommand{\ea}{\end{array}}
\def\labout{%
\setbox1=\hbox{$\sim$} \dimen1=\wd1                      
\hspace{0.3em}\rlap{\hbox to \dimen1{\hfil\raise.08em\hbox{$<$}\hfil}} 
\lower.37em\hbox{$\sim$}                                 
\raise.36em\hbox{}\hspace{0.3em}
}

\begin{document}
\draft
\title{\bf Giant phase-conjugate reflection with a normal 
mirror in front of an optical phase-conjugator}

\author{M. Blaauboer, D. Lenstra, and  A. Lodder}

\address{Faculteit Natuurkunde en Sterrenkunde, Vrije Universiteit,
         De Boelelaan 1081, 1081 HV Amsterdam, The Netherlands}

\date{\today}
\maketitle

\begin{abstract}
We theoretically study reflection of light by a phase-conjugating 
mirror preceded by a partially reflecting normal mirror. 
The presence of a suitably chosen normal mirror in front of the 
phase conjugator is found to greatly enhance the total 
phase-conjugate reflected power, even up to an order of magnitude.
Required conditions are that the phase-conjugating mirror itself
amplifies upon reflection and that constructive interference
of light in the region between the mirrors takes place.
We show that the phase-conjugate reflected power then 
exhibits a maximum as a function of the transmittance of 
the normal mirror. 
\end{abstract}

\pacs{PACS number(s): 42.65.Hw, 42.25.Bs, 42.68.Ay
       {\tt physics/9804003}}

\narrowtext

In recent years, the interesting role that phase conjugation can play
in mesoscopic physics has received considerable attention.
Phase conjugation is the general term for a process in which both 
the direction of propagation and
the overall phase factor of a wave function are reversed\cite{fisher83}.
A famous example is Andreev reflection\cite{andreev64}, the electron-to-hole
reflection at an interface between a normal metal and a superconductor
in a mesoscopic system. Mesoscopic means, by definition, that the dimensions
of the system are small enough for phase-coherence to be preserved
in the entire configuration, in this case the normal metal--superconductor
junction, but still much larger than the Fermi wavelength of the normal
metal. In this regime it is interesting to study the influence of 
Andreev reflection on transport properties such as the conductance of 
the sample. The basis
of a theoretical approach to this kind of transport problems was in fact 
laid as
early as 1957 by Landauer, long before the advent of mesoscopic physics.
He related the electrical conductance of a normal metal to its
quantum mechanical transmission matrix, the well-known Landauer 
formula\cite{landauer57,imry86}. This reflection/transmission matrix 
approach turned
out to be very useful in mesoscopic transport problems, also in
combination with phase conjugation. For example, Beenakker\cite{beenakker92} 
derived a "Landauer-type" formula for the conductance of a normal 
metal-superconductor junction, which allows for studying
both clean and disordered normal metals in the presence of Andreev
reflection.

This paper is concerned with the optical counterpart of Andreev reflection,
namely probe-to-conjugate reflection of light at a phase-conjugating 
mirror (PCM)\cite{fisher83}. The
dimensions of a PCM are much larger than those of a mesoscopic
superconducting system and allow for a classical treatment of the
transport properties of light. Despite this different scale, a PCM
system is mesoscopic in the sense that it is larger than the wavelength,
but smaller than the coherence length of the light reflecting at it.
Electronic and optical phase conjugation display an interesting
analogy\cite{lenstra90,vanhouten91} and also optical phase
conjugation can conveniently be described using a reflection/transmission 
formalism.
Recently this was used to study reflection of light at a phase-conjugating 
mirror behind a disordered optical medium in a waveguide\cite{paas97}.

In view of the time-reversal properties of a phase-conjugating
mirror, whereby accumulated phase-shifts are cancelled when the 
phase-conjugated wave travels back along the path of the original incoming
wave, it is
intruiging to combine it with a normal reflector. By normal reflector
we mean a scattering region in which only specular reflections
take place, eg. the disordered medium in\cite{paas97}. Here
we consider the simplest possible normal reflector, a partially transmitting
normal mirror. We analyze the phase-conjugate reflected power
for plane wave illumination of a configuration consisting of a 
phase-conjugating mirror
preceded by a partially reflecting normal mirror. A lot of work
has been done on this kind of resonator structures involving combinations
of normal mirrors and phase-conjugating mirrors\cite{previous}. Here we 
describe an interesting effect in a normal mirror--phase-conjugating 
mirror arrangement, which to our knowledge has been unnoticed so far.
When the PCM is operating such that it amplifies the incoming light
upon phase-conjugate reflection and multiply reflected waves
in the region between the mirrors interfere constructively,
we find a dramatic
enhancement of the phase-conjugate reflected intensity compared with 
that at the same PCM alone. The phase-conjugate reflectance, defined
as the reflected power divided by the incident power, is maximal 
for a suitably chosen value of the transmittance of the normal mirror
and reaches $\sim 10$ times its value at the PCM alone.

The phase-conjugating mirror consists of a cell filled with
an optical medium with a large third-order susceptibility $\chi^{(3)}$. 
The medium is pumped by two intense counterpropagating laser beams of 
frequency $\omega_{0}$. When a weak probe beam of frequency $\omega_{0} 
+ \delta$ is incident on the material, a fourth beam will be generated 
due to the nonlinear polarization of the medium (four-wave mixing). 
This so-called 
conjugate wave propagates with frequency $\omega_{0} - \delta$ in the 
opposite direction as the probe beam\cite{fisher83}.
In the medium, the coupling between probe and conjugate waves is
described by the equations\cite{lenstra90}
\be
\left( \begin{array}{cc}
- \frac{c^2}{2\omega_{0}} \frac{\partial^2}{\partial x^2} - 
\frac{\omega_{0}}{2}  & - \gamma \vspace{0.3cm} \\
\gamma^{*} &  \frac{c^2}{2\omega_{0}} \frac{\partial^2}{\partial x^2}
+ \frac{\omega_{0}}{2} \end{array} \right)
\left( \begin{array}{l}
{\cal E}_{p}(x) \vspace{0.3cm} \\ {\cal E}_{c}^{*}(x)
\end{array} \right) = \delta
\left( \begin{array}{l}
{\cal E}_{p}(x) \vspace{0.3cm} \\ {\cal E}_{c}^{*}(x)
\end{array} \right).
\label{eq:SEFL}
\ee
Here ${\cal E}_{p (c)}(x)$ denotes the complex amplitude of the
probe (conjugate) field and $\gamma \equiv \gamma_{0} e^{i\phi}$ 
is the pumping induced coupling strength between the two fields.
Outside the medium $\gamma=0$ and (\ref{eq:SEFL}) reduces to
two uncoupled equations for the probe and conjugate waves.

\begin{figure}
\centerline{\epsfig{figure=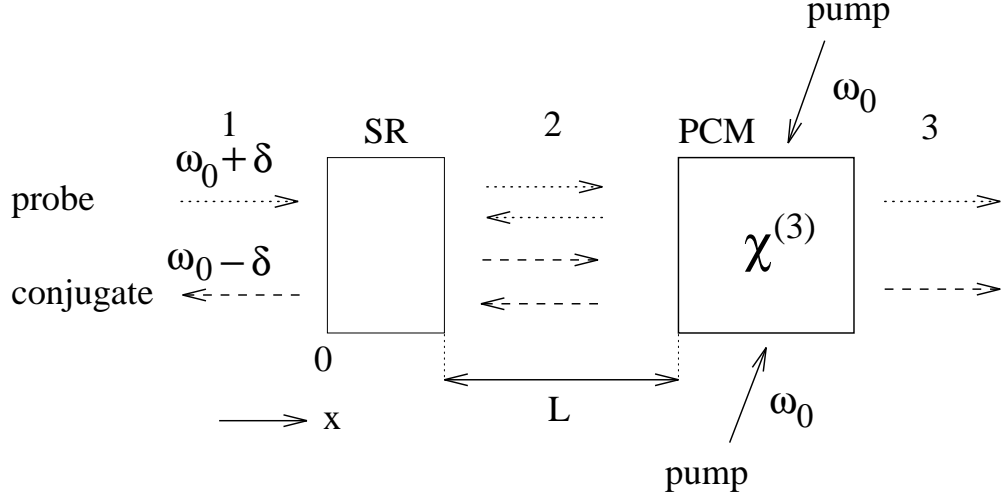,width=0.8\hsize}\vspace{0.5cm}}
\caption{Reflection and transmission in one dimension at
a scattering region (SR) followed by a phase-conjugating mirror (PCM). 
Dotted (dashed) arrows denote probe (conjugate) waves. 
}
\label{fig:system}
\end{figure}

\begin{figure}
\centerline{\epsfig{figure=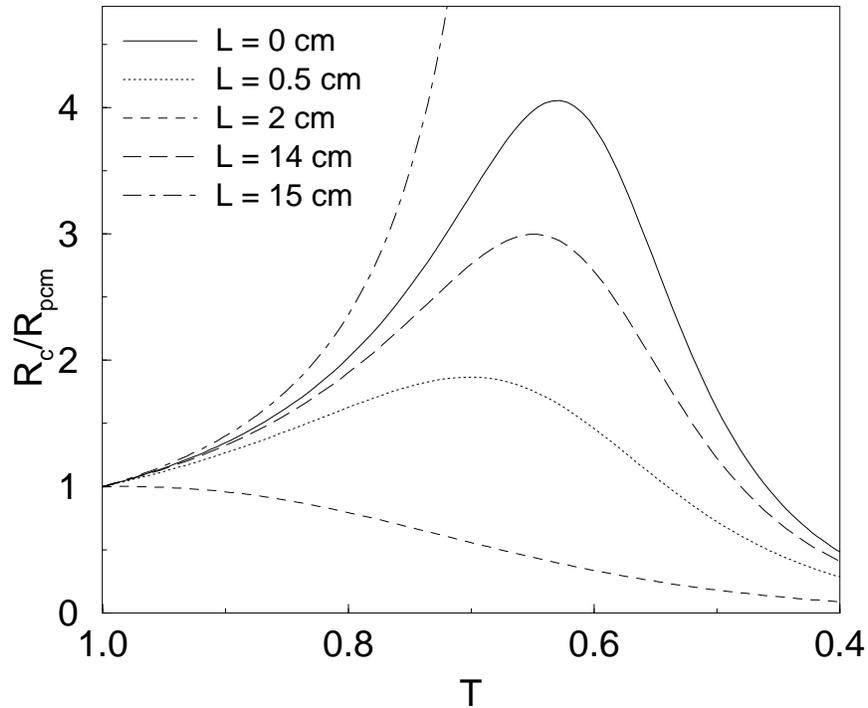,width=0.8\hsize}}
\caption{Phase-conjugate reflectance at a normal mirror
with transmittance $T$ followed by a phase-conjugating mirror 
for $L_{pcm} = 1\, cm$, $\delta=0.1\, [c/L_{pcm}]$, 
$\gamma_{0}=1\, [c/L_{pcm}]$.
The phase-conjugate reflectance of the PCM alone is
$R_{pcm}=2.4$
} 
\label{fig:Rc1}
\end{figure}

The one-dimensional system we consider is schematically depicted
in Fig.~\ref{fig:system}. A phase-conjugating mirror is preceded at 
distance $L$ by, in general, an elastic scattering region (SR) which 
does not couple probe and conjugate waves.  
We are interested in the phase-conjugate reflected intensity 
at $x=0$ if a probe beam is incident on the region from the left. 
This can be straightforwardly calculated using a 
scattering matrix formalism.
Let ${\cal E}_{p(c), 1(2)}^{+}$ denote the electric field amplitude
of a probe (conjugate) wave travelling to the right (left) in region 1 (2).
We then have \vspace{0.3cm}
\be
\left( \ba{l}
{\cal E}_{p,3}^{+}(x) \vspace{0.13cm} \\
{\cal E}_{p,3}^{-}(x) \vspace{0.13cm} \\
{\cal E}_{c,3}^{-}(x) \vspace{0.13cm}\\
{\cal E}_{c,3}^{+}(x)
\ea \right) \ = \
S_{pcm}\ U(x)\ S(x)\ 
\left( \ba{l}
{\cal E}_{p,1}^{+}(x) \vspace{0.13cm}\\
{\cal E}_{p,1}^{-}(x) \vspace{0.13cm}\\
{\cal E}_{c,1}^{-}(x) \vspace{0.13cm}\\
{\cal E}_{c,1}^{+}(x)
\ea \right),
\vspace{0.3cm}
\label{eq:matrixeq}
\ee
with $S_{pcm}$ the scattering matrix of the PCM, $S$ that of the scattering
region in front and $U$ the transfer matrix in region 2.
From the solution of the matrix equation (\ref{eq:matrixeq}) for known $S$ 
and $S_{pcm}$ the phase-conjugate reflected intensity 
$R_{c} \equiv |{\cal E}_{c,1}^{+}(0)|^2$ at $x=0$ is obtained.
In general $R_{c}$ can be written as
\be
R_{c} = \frac{T_{p}\, T_{c}\, R_{pcm}}{1 + (1- T_{p})(1 - T_{c}) R_{pcm}^2 
- 2 R_{pcm} \left[ \mbox{\rm Re}(r_{p,2} \cdot r_{c,2}) \cos \phi - 
\mbox{\rm Im}(r_{p,2} \cdot r_{c,2}) \sin \phi \right]}.
\label{eq:conj1}
\ee
Here $T_{p (c)}$ denotes the transmittance of probe (conjugate) waves 
through the scattering region and Re\,($r_{p,2}$) is the real part of
the amplitude of probe-to-probe reflection at SR in region 2, etc. 
$\phi$ is the phase accumulated during multiple reflections in region 2
\cite{bound},
\be
\phi = 4\frac{\delta}{c}L + 2 
\arctan(\frac{\delta}{\sqrt{\delta^2 + \gamma_{0}^2}}\tan(\beta L_{pcm}))
\label{eq:phi}
\ee
and $R_{pcm}$ is the phase-conjugate reflected power at the PCM
alone\cite{fisher83},
\be
R_{pcm}  =  \frac{\sin ^2 (\beta L_{pcm})}{\cos^2 (\beta L_{pcm}) + 
(\frac{\delta}{\gamma_{0}})^2}, 
\ee
with $\beta = \sqrt{\delta^2 + \gamma_{0}^2}/c$ and $\delta$ the detuning
frequency between probe and pump waves, $\delta \ll \omega_{0}$. 
If the scattering region consists of a single partially
transmitting normal mirror, (\ref{eq:conj1})
reduces to 
\begin{eqnarray}
R_{c} & = & \frac{T^2 R_{pcm}}{1 + (1- T)^2 R_{pcm}^2 
- 2 (1-T) R_{pcm} \cos \phi} \nonumber \\
& = & \frac{T^2 R_{pcm}}{\left[ 1 - (1-T) R_{pcm} \right]^2 +
2(1-T)R_{pcm}(1-\cos \phi)},
\label{eq:conj2}
\end{eqnarray}
with $T \in [0,1]$ the transmittance of the normal mirror at frequency 
$\omega_{0}$\cite{trans}.
Fig.~\ref{fig:Rc1} shows $R_{c}$ as a function of T for various values
of the intermirror distance $L$.
We see that the three middle curves in Fig.~\ref{fig:Rc1}
display a maximum as a function of $T$. More precisely,
the maximum in the phase-conjugate reflected intensity
occurs if $R_{pcm} \cos \phi \geq 1$  and for normal
mirror transmittance 
\be
T_{max} = \frac{1 - 2 R_{pcm} \cos \phi + R_{pcm}^2}{R_{pcm} (R_{pcm}
-\cos \phi)}.
\ee
$R_{c}$ then reaches a value which amounts to several times the 
phase-conjugate reflected intensity at a PCM alone.

Placing a normal mirror in front of a phase-conjugating mirror thus 
greatly enhances the phase-conjugate reflected intensity, as long as 
the PCM
acts as an amplifier ($R_{pcm}>1$) and the phase $\phi$ is such that
$\cos \phi \sim 1$.
The maximum occurs because of two competing effects: on the one hand,
the normal mirror causes direct back-reflection of part of the incoming
probe wave, so that less light reaches the PCM. On the other hand it
opens the possibility for multiple reflections in the region between 
the two mirrors. For large normal mirror transmittance the latter effect 
is dominant,
because of the gain in intensity for each reflection at the PCM and
the constructive interference of light in the resonator ($\phi \sim 2\pi n$). 
As $T$ becomes smaller, however, the increasing loss of light through 
backscattering without phase-conjugation causes $R_{c}$ to drop again.
The amplifying property of the PCM is essential for obtaining
the maximum. This can easily be seen from (\ref{eq:conj2}) which shows that
$R_{c} < R_{pcm}$ for $R_{pcm} < 1$\cite{larger}.
The constructive interference
of multiply reflected waves is also essential for obtaining the maximum.
Fig.~\ref{fig:Rc1} shows that for $L=2\, cm$ (or, equivalently, $\phi 
\approx \pi/3$), $R_{c}$ always decreases
with $T$, even though the PCM still acts as an amplifier. 
The advantage of additional multiple reflections, as the 
transmittance of the normal mirror becomes less, then cannot overcome
the disadvantage of less primary light arriving at the PCM.
Note that a finite distance $L$ is not necessary for obtaining
the maximum. The curve for $L=0$ shows that a suitably chosen coating
on the PCM also gives rise to enhancement of $R_{c}$. In this case
the effect is very sensitive to the detuning frequency $\delta$ which, as seen 
from (\ref{eq:conj2}) with $L=0$, has to satisfy the condition
$\delta \ll \gamma_{0}$ in order for $\phi$ to be close to zero.

From (\ref{eq:conj2}) we see that $R_{c}$ becomes infinite if
$\cos \phi=1$ and $T = 1 - 1/R_{pcm}$. This happens in the upper curve
of Fig.~\ref{fig:Rc1}, for $L=15\, cm$. It is interesting to compare
this with the work of Paasschens {\it et al.} who obtained
qualitatively the same divergent behavior in a more complex 
system\cite{paas97}. They studied
reflection of light at a two-dimensional random medium consisting 
of dielectric rods backed by a phase-conjugating mirror. For diffusive 
illumination of the disordered medium and for $\delta \gg \tau_{dwell}^{-1}$,
with $\tau_{dwell}$ the dwell time of a photon in the medium, 
interference effects are shown to be negligible. Disregarding angular 
correlations
between multiply reflected waves in the region between the random
medium and the PCM then leads to a phase-conjugate reflected power
of
\be
R_{c} = \frac{T^2 R_{pcm}}{1 - (1-T)^2 R_{pcm}^2}.
\label{eq:conj3}
\ee
The same expression is obtained when averaging (\ref{eq:conj2}) over 
the phase $\phi$\cite{fase}. (\ref{eq:conj3}) again shows the divergence 
of $R_{c}$ for $T = 1 - 1/R_{pcm}$. As was explained in
\cite{paas97}, this is due to some light becoming trapped in the intermirror 
region as $T$ decreases. Repetitive reflections
at the PCM, provided $R_{pcm} >1$, then lead to the enormous 
increase in the phase-conjugate reflectance.
Since $R_{c}$ is limited by the intensity of the pump beams, one now needs
to take into account the effects of pump depletion and the analysis
breaks down close to and beyond $T=1 - 1/R_{pcm}$.

\begin{figure}
\centerline{\epsfig{figure=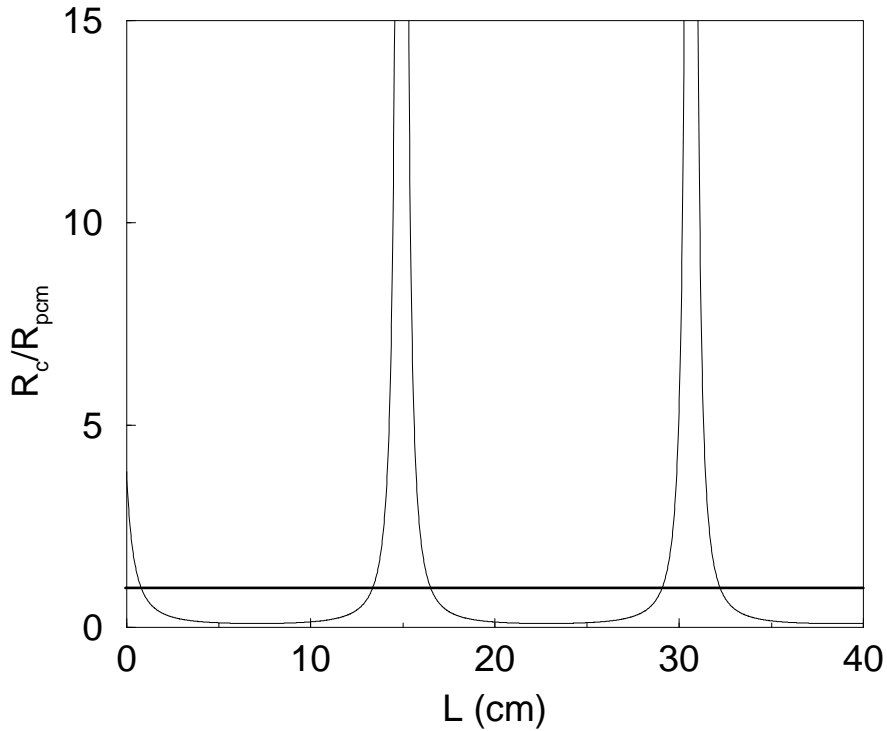,width=0.8\hsize}}
\caption{Phase-conjugate reflectance at a normal mirror
with transmittance $T=0.6$ followed by a phase-conjugating mirror. 
Parameters used are $L_{pcm} = 1\, cm$, $\delta=0.1\, [c/L_{pcm}]$ and
$\gamma_{0}=1\, [c/L_{pcm}]$.
The horizontal line marks the
phase-conjugate reflectance at the PCM alone, 
$R_{pcm}=2.4$.
} 
\label{fig:Rc5}
\end{figure}

The same phenomenon occurs in our one-dimensional system for certain
values of $L$\cite{difference}. This is illustrated in Fig.~\ref{fig:Rc5}, 
which shows 
the phase-conjugate reflected intensity as a function of the intermirror
distance $L$ for a normal mirror with transmittance of 60 \%. The horizontal
line in the figure indicates the reflectance of the PCM alone, $R_{pcm}$.
It is clear that the presence of  the normal mirror at a suitably
chosen distance $L$, such that $\phi$ is close to a multiple of $2\pi$,
but sufficiently far away from it to avoid the effects of pump depletion,
gives a giant enhancement of $R_{c}$ with respect to $R_{pcm}$.

In conclusion, we have theoretically analyzed the phase-conjugate reflected
power at a normal mirror followed by a phase-conjugating mirror.
For plane wave illumination of this mirror arrangement, and taking advantage
of multiple reflection effects in the resonator region between the mirrors, 
a giant 
enhancement of the phase-conjugate reflectance is predicted with respect to that
of the PCM alone. Necessary conditions are that reflection
at the phase-conjugating mirror is accompagnied by amplification, and that
constructive interference of waves occurs in front of the PCM.
If either condition is not satisfied, the presence of the normal mirror 
only decreases the phase-conjugate reflectance. This also happens in 
the analogous electronic
configuration, a normal metal-superconductor interface preceded by a 
normal scattering region: since Andreev reflection occurs with at most unity 
amplitude, the hole reflectance decreases with increasing
normal scattering\cite{beenakker94}. 
 
In previous work on resonators involving a normal mirror and a 
phase-conjugating mirror\cite{previous} various ways of output power
enhancement of the resonator were found. For example, Pepper 
{\it et al.}\cite{pepper78}
reported amplified phase-conjugate reflection using degenerate four-wave
mixing ($\delta = 0$) and a fully reflecting normal mirror in combination
with a PCM. Under suitably chosen operating conditions of the PCM they 
observed a finite conjugate output signal in the absence of any input probe
signal, the output being directly generated by the pump beams\cite{self}. 
Feinberg {\it et al.}\cite{feinberg81} obtained similar self-oscillatory
behavior using a perfectly reflecting normal mirror behind a PCM.

Our configuration differs from these resonators, since it gives
the phase-conjugate response to an input probe signal passed
through a partially transmitting normal mirror. Enhancement
of this response occurs both for frequencies $\delta = 0$ 
(degenerate four-wave mixing) and $\delta \neq 0$ (non-degenerate
four-wave mixing), and in a broad range of intermirror
distances $L$. 
This allows for frequency discrimination
between probe and conjugate output waves and regularization
of the enhancement by choosing an appropriate value of $L$ for
fixed $\delta$, or vice versa.

Finally, as an example, consider a realistic phase-conjugating mirror
of length $L_{pcm} = 1\, cm$, with a 
coupling strength $\gamma_{0} = 10^{10} s^{-1}$ and
illuminated by monochromatic light of frequency $\delta \sim 
10^9 s^{-1}$ detuned from the pump frequency. 
The phase-conjugate reflectance
$R_{pcm} \sim 2$, which is well within experimental reach\cite{lanzerotti96}.
We predict that placing a $\approx 60 \%$ transmitting normal mirror
at a distance of $L \approx 14\, cm$ in front of this PCM increases
the phase-conjugate reflected power by an order of magnitude
and hope that this will present a challenge to experimentalists.

Part of this work was supported by the "Stichting voor Fundamenteel
Onderzoek der Materie" (FOM) which is part of the "Nederlandse Organisatie 
voor Wetenschappelijk Onderzoek" (NWO).


\begin{references}
\bibitem{bound}    (\ref{eq:phi}) is the condition for the formation
	        of a bound state in a normal mirror--phase-conjugating
	          mirror resonator\cite{vanhouten91}.
	          Due to the coupling of probe and conjugate waves by the
	          PCM, a complete roundtrip requires four passages
	          through the resonator region (first term on right-hand
	          side of (\ref{eq:phi})) and two reflections at
	          the PCM (second term).
\bibitem{trans}   We neglect the difference in normal-mirror transmission
		  of probe and conjugate waves here, since 
		  $\delta \ll \omega_{0}$.
\bibitem{larger} If reflection at the PCM is not accompagnied by
		amplification, enhancement of the phase-conjugate reflectance
		can be achieved by placing a normal mirror \underline{behind}
		the phase-conjugating mirror and illuminating the PCM 
		with probe light in front.
		The total phase-conjugate reflectance at the PCM then
		consists of a part resulting from direct phase-conjugate
		reflection at the PCM plus an extra part as a result of 
		multiple reflections in the region between phase-conjugating
		mirror and normal mirror.
\bibitem{fase}     In performing this average we in fact consider a 
		 somewhat hypothetical "two-dimensional"
	         system consisting of a sum of $N$ one-dimensional channels
	         (eg. fibres), each containing a normal mirror, in front
	         of the PCM. If the positions of the normal mirrors
	         in each of these fibres are random, the accumulated
	         phases $\phi_{n}, n=1 \dots N$ will vary uniformly 
	         over integral multiples of $2\pi$ for $N\gg 1$, and one 
	         can average over them.
\bibitem{difference} Notice that the behavior of the phase-conjugate
	        reflectance \underline{before} the enormous increase at
	        $T=1 - 1/R_{pcm}$ is different in the one-dimensional and
	        two-dimensional situation. In one dimension $R_{c}$ 
	        immediately increases as $T$ becomes less than 1 (upper curve
	        in Fig.~\ref{fig:Rc1}). In two dimensions it first decreases
	        and displays a minimum at $T=1 - 1/R_{pcm}^2$\cite{paas97} 
	        as a result of the diffusive illumination and subsequent
	        angular, or phase-shift average.
\bibitem{self} The condition for this so-called self-oscillating
	         behavior is the same  for which $R_{c} \to \infty$ in	         
	         (\ref{eq:conj2}) with $\delta=0$, 
	         $R_{pcm} = 1/1-T$\cite{pepper78}.
\end{references}
\end{document}